\documentclass[aps,prl,nopacs,twocolumn,preprintnumbers,notitlepage,amsmath,amssymb,superscriptaddress]{revtex4-2}
\usepackage[colorlinks=true,allcolors=blue]{hyperref} 
\usepackage{amssymb}
\usepackage{amsmath}
\usepackage{color}
\usepackage[pdftex]{graphicx}
\usepackage{bm}  
\usepackage{graphicx}
\usepackage{amscd}
\usepackage{epsfig}
\usepackage{epstopdf}
\usepackage{enumerate}
\usepackage{mathtools}
\usepackage{notes2bib}

\newcommand{\beq}{\begin{equation}}
\newcommand{\eeq}{\end{equation}}
\newcommand{\beqa}{\begin{eqnarray}}
\newcommand{\eeqa}{\end{eqnarray}}
\newcommand{\bem}{\begin{math}}
\newcommand{\eem}{\end{math}}

\newcommand{\bfr}{{\bm r}}

\newcommand{\bfq}{{\bm q}}

\newcommand{\bff}{{\bm f}}
\newcommand{\bfv}{{\bm v}}

\newcommand{\bfE}{{\bm E}}

\newcommand{\bnabla}{{\bm \nabla}}

\newcommand{\aver}[1]{\left\langle {#1}\right\rangle}
\newcommand{\lae}{\lambda_{\rm e}}
\newcommand{\bee}{\hat{\bm e}}
\newcommand{\ee}{\hat{e}}
\newcommand{\bqhat}{\hat{\bm{q}}}
\newcommand{\qhat}{\hat{q}}

\begin{document}

\title{Anomalous Diffusion in Driven Electrolytes due to Hydrodynamic Fluctuations}

\author{Ramin Golestanian}
\email{ramin.golestanian@ds.mpg.de}
\affiliation{Max Planck Institute for Dynamics and Self-Organization (MPI-DS), 37077 G\"ottingen, Germany}
\affiliation{Rudolf Peierls Centre for Theoretical Physics, University of Oxford, Oxford OX1 3PU, United Kingdom}

\date{\today}

\begin{abstract}
The stochastic dynamics of tracers arising from hydrodynamic fluctuations in a driven electrolyte is studied using a self-consistent field-theory framework in all dimensions. A plethora of scaling behaviour that includes two distinct regimes of anomalous diffusion is found, and the crossovers between them are characterized in terms of the different tuning parameters. A short-time ballistic regime is found to be accessible beyond two dimensions, whereas a long-time diffusive regime is found to be present only at four dimensions and above. The results showcase how long-ranged hydrodynamic interactions can dominate the dynamics of non-equilibrium steady states in ionic suspensions and produce strong fluctuations despite the presence of Debye screening.  
\end{abstract}

\maketitle

\textit{Introduction.---}Since the pioneering works of Faraday \cite{Faraday1839} and Nernst \cite{Nernst1889}, driven electrolytes have been at the forefront of many scientific and technological developments to date, most notably, in energy storage and conversion applications. Moreover, evolution has selected them as the physical medium of operation for the most sophisticated signal processing and cognitive biological systems. In recent years, understanding and controlling the physical characteristics of driven electrolytes at the nano-scale have enabled the development of molecular sequencing and sensing technologies \cite{Ying2022,Ritmejeris2024}, \textit{de novo} strategies to build synthetic nanoscale motors \cite{Pumm2022,Shi2022,Shi2024a}, as well as theoretical proposals to explain complex biological processes such as the gating of wet ion channels via an intrinsically non-equilibrium mechanism \cite{Bonthuis2014} and neuro-morphic computing using solid-state nano-pores \cite{Siwy2002,Stuhlmller2025}. On the other hand, driven electrolytes provide a playground for the study of non-equilibrium steady states that exhibit a rich variety of interesting physical behaviour \cite{Mahdisoltani2023}. Recent examples include the emergence of long-range correlations and fluctuation-induced forces between boundaries immersed in driven electrolytes \cite{Mahdisoltani2021b,mahdisoltani2021,Du2025,Bonneau2025} and related generalizations \cite{Grabsch2025}, as well as probes of the so-called $1/f$ noise in ionic currents in nano-pores \cite{zorkot2016a,zorkot2016,zorkot2018nanopore,Marbach2021,Gubbiotti31122022}. 

\begin{figure}[b]
\centering
\includegraphics[width=0.99\columnwidth]{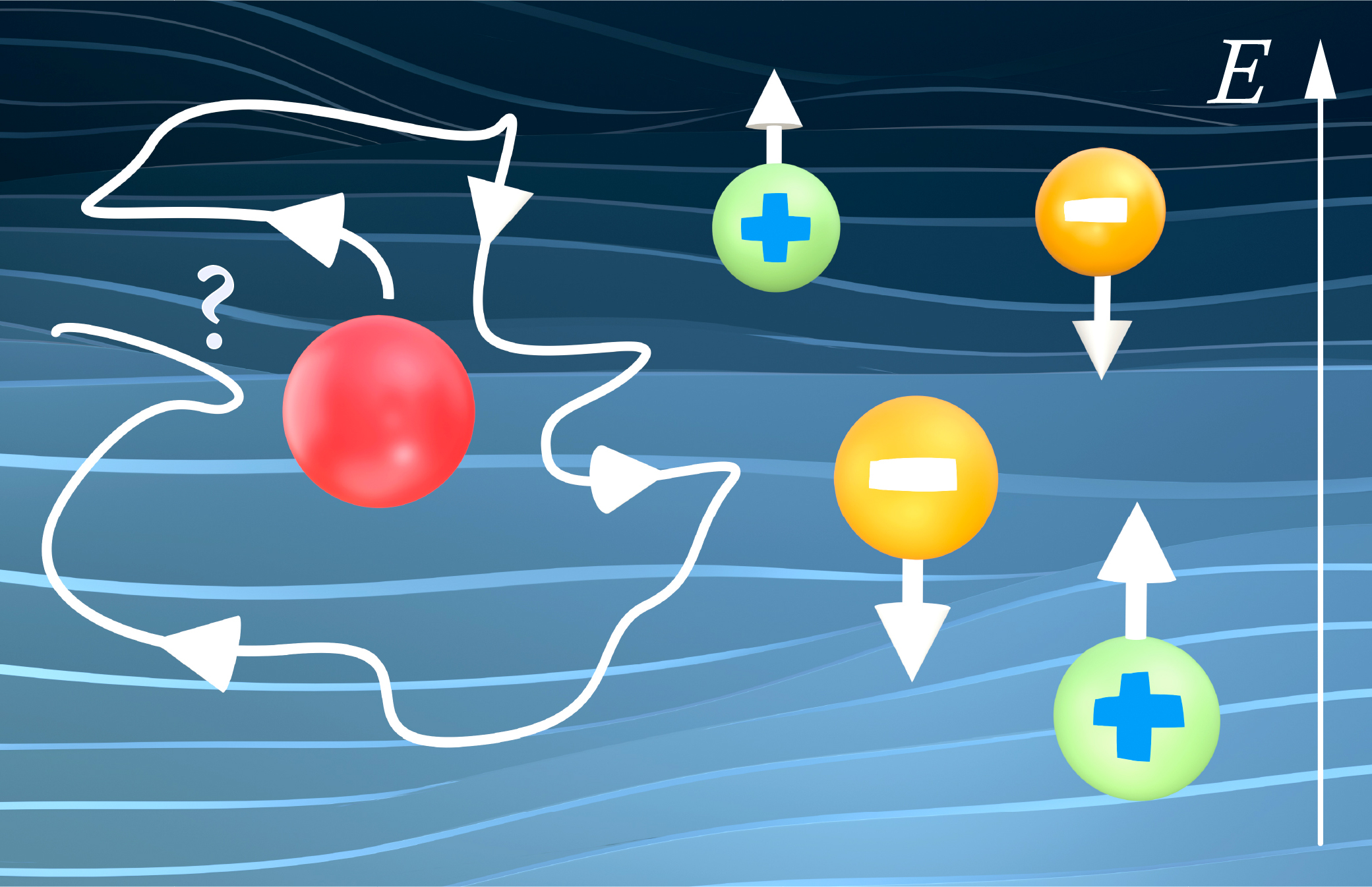}
\caption{An electrolyte driven by an applied electric field $E$ encompasses body forces exerted on the positive and negative ions in the fluid, which possess a dipolar nature due to the overall charge neutrality. These fluctuating body forces lead to the generation of long-ranged flow fields that can be measured through the stochastic trajectories of tracer particles.}\label{fig:scheme}
\end{figure}

The role of the solvent on the stochastic dynamics of mobile ions and tracers in electrolytes has so far received relatively little attention. Hydrodynamic interactions induce long-time tails on tracers in neutral solvents \cite{Alder1967}, and introduce strong large-scale fluctuations in the phenomenon of sedimentation, where body forces that scale with the system size are exerted externally on the colloidal particles in the suspension \cite{Sriram1998,Miguel2001}. In electrolytes, the body forces appear in pairs due to the overall neutrality of the suspension, and this might introduce additional subtleties, not unrelated to the force-free nature of interfacial transport processes that are relevant to active matter systems \cite{Anderson1982,Golestanian2019phoretic}. In this Letter, the aim is to study the effect of hydrodynamic fluctuations that arise from the non-equilibrium electrical force dipoles that stir up a suspension in an electrolyte driven by an external electric field; see Fig. \ref{fig:scheme}. Using a self-consistent field-theory framework, the stochastic dynamics of tracers in a driven electrolyte is studied as driven by the fluctuations in the hydrodynamic flow field, covering all dimensions.

\begin{figure*}[t]
\centering
\includegraphics[width=0.5\columnwidth]{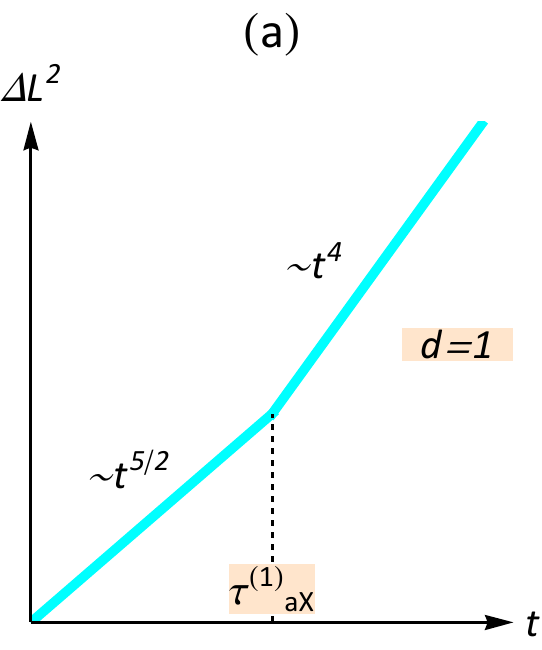}
\includegraphics[width=0.5\columnwidth]{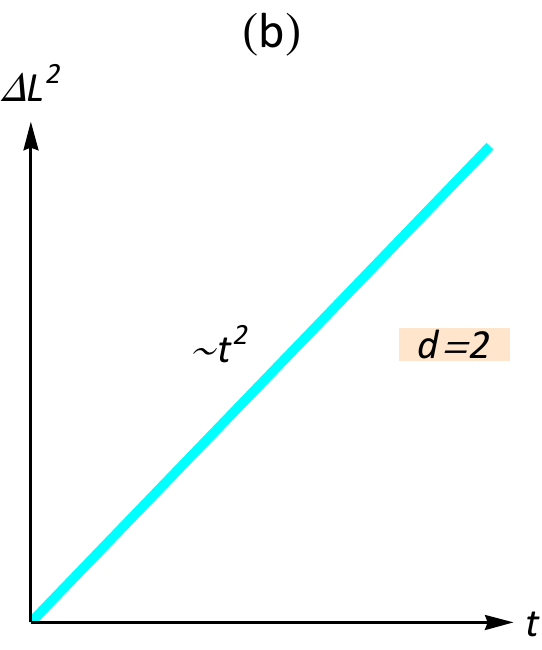}
\includegraphics[width=0.5\columnwidth]{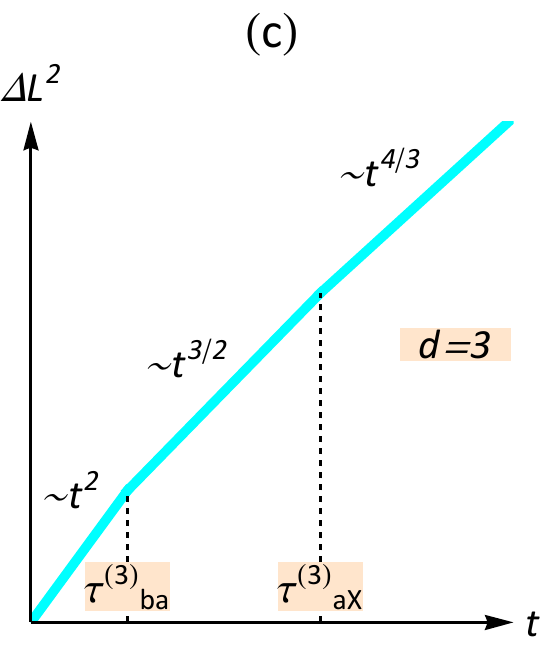}
\includegraphics[width=0.5\columnwidth]{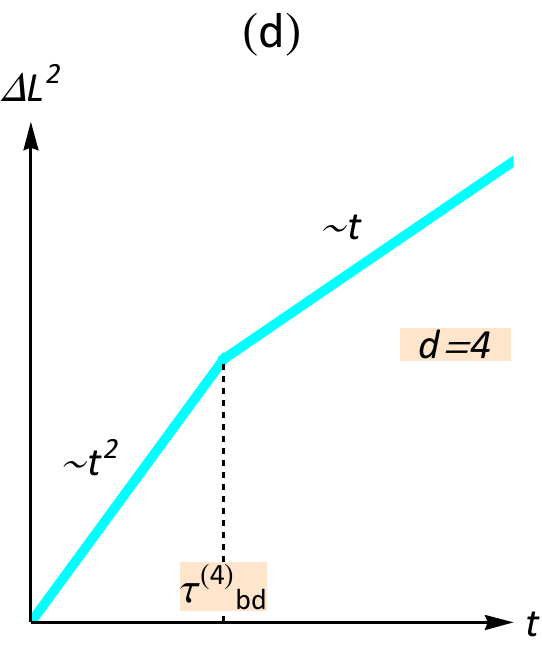}
\caption{Summary of the different regimes of the dynamics in different dimensions. (a) In $d=1$, there is a crossover between two super-ballistic anomalous regimes occurring at $\tau^{(1)}_{\textrm{a}\times}=C_0^{2/3} D^{1/3} \lae^{-4/3}$. (b) In $d=2$, there is only one ballistic regime observed at all times. (c) In $d=3$, the dynamics shows a crossover from ballistic to a first anomalous regime at the time-scale  $\tau^{(3)}_{\textrm{b}\textrm{a}}=a^2 D^{-1}$, followed by another crossover to a second anomalous regime at the time-scale $\tau^{(3)}_{\textrm{a}\times}=C_0^{2} D^{3} \lae^{-4}$. (d) In $d=4$, there is a crossover from a ballistic regime to a diffusive regime at the time-scale $\tau^{(4)}_{\textrm{b}\textrm{d}}=C_0^{1/2} \lae^{-1} a^2 \sqrt{\ln(L/a)}$. In $d>4$, the same crossover occurs from ballistic to diffusive behaviour at the time-scale $\tau^{(d)}_{\textrm{b}\textrm{d}}=C_0^{1/2} \lae^{-1} a^{d/2}$.}\label{fig:MSD}
\end{figure*}

We now present a summary of the main results on the way the mean-squared displacement (MSD) of tracers in the electrolyte denoted as $\Delta L^2(t)$ depends on time $t$ in dimension $d$, covering all dimensions (see Fig.~\ref{fig:MSD}). The electrolyte comprises a suspension of monovalent ions with charges $\pm Q$ and mean concentration $C_0$ in a solvent with viscosity $\eta$ and dielectric constant $\epsilon$. The diffusion coefficient $D$ of the individual ions is taken to be equal for both positive and negative species. An electric field $\bfE=E \bee$ is applied globally on the system (see Fig.~\ref{fig:scheme}). Electrostatic effects are subject to Debye screening due to the thermal fluctuations of the ions in the medium with inverse temperature scale $\beta=1/(k_{\rm B} T)$, and the Debye screening length $ \kappa^{-1} $ is defined via $\kappa^2 = 2 S_d C_0 \beta Q^2/\epsilon$ in $d$ dimensions (with $S_d={2\pi^{d/2}}/{\Gamma ({d}/{2})}$, e.g. $S_3=4\pi$). An important inverse time-scale in the system is given by $\lae=\epsilon E^2/(2 S_d\eta)$, which represents the typical scale of the shear strain rate in the electrolyte suspension arising from the Maxwell stress associated with the electric field. A microscopic length scale $a$, which represents the size of the tracers or $\kappa^{-1}$ (whichever is larger), and the system size $L$ are used in the regularization of the theoretical framework. The system exhibits ballistic, anomalous, and diffusive regimes, as characterized by the time dependence of the MSD, defined as $\Delta L^2 \equiv \aver{[\bfr(t)-\bfr(0)]^2}$, where $\bfr(t)$ describes the stochastic trajectory of the tracer particle. In $d=1$, the dynamics at short times obeys $\Delta L^2 \sim t^{5/2}$ for $0 < t < \tau^{(1)}_{\textrm{a}\times}$, which is followed by a crossover to $\Delta L^2 \sim t^{4}$ for $t > \tau^{(1)}_{\textrm{a}\times}$. The crossover time scale between the two super-ballistic anomalous regimes is found as $\tau^{(1)}_{\textrm{a}\times}=C_0^{2/3} D^{1/3} \lae^{-4/3}$. In $d=2$, only one ballistic regime is observed with $\Delta L^2 \sim t^{2}$ at all time scales. In $d=3$, the dynamics exhibits ballistic behaviour with $\Delta L^2 \sim t^{2}$ at short times $0<t<\tau^{(3)}_{\textrm{b}\textrm{a}}$, followed by a crossover beyond the time-scale $\tau^{(3)}_{\textrm{b}\textrm{a}}=a^2 D^{-1}$ to a first anomalous regime described by $\Delta L^2 \sim t^{3/2}$ for $\tau^{(3)}_{\textrm{b}\textrm{a}} < t < \tau^{(3)}_{\textrm{a}\times}$, which is subsequently followed by a crossover to a second anomalous regime that obeys $\Delta L^2 \sim t^{4/3}$ for $t > \tau^{(3)}_{\textrm{a}\times}$. The crossover time-scale between the two anomalous regimes is obtained as $\tau^{(3)}_{\textrm{a}\times}=C_0^{2} D^{3} \lae^{-4}$. In $d=4$, the dynamics starts with ballistic behaviour $\Delta L^2 \sim t^{2}$ at short times $0<t<\tau^{(4)}_{\textrm{b}\textrm{d}}$, and exhibits a crossover at the time-scale $\tau^{(4)}_{\textrm{b}\textrm{d}}=C_0^{1/2} \lae^{-1} a^2 \sqrt{\ln(L/a)}$ to a diffusive regime $\Delta L^2 \sim t$ for $t > \tau^{(4)}_{\textrm{b}\textrm{d}}$. In $d>4$, the same crossover from ballistic to diffusive behaviour is observed at the time-scale $\tau^{(d)}_{\textrm{b}\textrm{d}}=C_0^{1/2} \lae^{-1} a^{d/2}$. These results are presented in Fig.~\ref{fig:MSD}.

\textit{Theoretical framework.---}The dynamics of the ions is probed by using the stochastic concentrations of the two charged species $ C^\pm(\bm{r},t)$ that give rise to a stochastic electrostatic potential field $\phi(\bm{r},t)$ through the Poisson equation $ -\bnabla^2 \phi = S_d  Q( C^+ - C^-)/\epsilon$, which is written in Gaussian units in $d$ dimensions. Using the Dean--Kawasaki approach \cite{dean96langevin,kawasaki1994}, the dynamics of $C^\pm$ can be described by continuity equations, $ \partial_t C^\pm + \bnabla\cdot \bm{J}^\pm = 0$, where the stochastic currents are given as $\bm{J}^{\pm} = C^\pm \bm{v} - D \bnabla C^\pm \pm D \beta Q C^\pm \left(  - \bnabla\phi + \bm{E} \right) - \sqrt{2 D C^\pm} \, \bm{\eta}^\pm$. Here, $\bm{v}(\bm{r},t)$ describes the fluid flow and $\bm{\eta}^\pm$ are uncorrelated Gaussian white noise fields characterized by zero mean and $\langle \eta_i^\pm(\bm{r},t) \eta_j^\pm(\bm{r}',t') \rangle =\delta_{ij} \delta^d(\bm{r}-\bm{r}') \delta(t-t')$. In what follows, averaging will be performed with respect to the different realizations of this noise, e.g. $\aver{C^\pm}=C_0$, and it is convenient to write $C^\pm = C_0 + \delta C^\pm$ where the density fluctuation have zero mean, namely, $\aver{ \delta C^\pm}=0$.

The velocity field $\bfv(\bfr,t)$ in the background medium is subject to the incompressibility condition $\bnabla \cdot \bfv=0$ as well as momentum conservation, which is enforced via the Stokes equation $-\eta \bnabla^2 {\bm v}=-\bnabla p+\bff$ that describes local and instantaneous stochastic force balance. Here, $p(\bfr,t)$ is the pressure field and $\bff(\bfr,t)=Q \bfE \rho(\bfr,t)$ represents the body-force density experienced by the ions due to the electric field and non-vanishing stochastic charge density $\rho=(C^+ - C^-)$ (defined in units of $Q$). The Stokes equation can be solved exactly in terms of the charge density. The solution can be formally represented as
\beq
\bfv(\bfr,t)=\frac{QE}{\eta} \left(\frac{1}{-\bnabla^2}\right) \left(\bee-\frac{(\bee \cdot \bnabla)\bnabla}{\bnabla^2}\right) \rho,\label{eq:v-solution-formal}
\eeq
and the result can be used to derive an expression for the velocity fluctuations of the fluid medium in Fourier space as follows
\beqa
\aver{v_i(\bfq,\omega) v_j(-\bfq,-\omega)}&=&\left[\ee_i-\qhat_i (\bqhat\cdot \bee)\right] \left[\ee_j-\qhat_j (\bqhat\cdot \bee)\right] \nonumber \\
&\times&\frac{Q^2E^2}{\eta^2 q^4} \aver{|\rho(\bfq,\omega)|^2}.\label{eq:velo-fluc-1}
\eeqa
In the limit of high concentrations corresponding to strong electrolytes, the stochastic densities are relatively weak $\delta C^\pm \ll C_0$, and therefore the stochastic density equations can be expanded around the background mean. The total number density fluctuations $c (\bm{r},t) = \delta C^+  +  \delta C^-$ and the charge density fluctuations $\rho (\bm{r},t) = \delta C^+  -  \delta C^-$ satisfy the following stochastic equations
\beq 
    \partial_t c + \bfv \cdot \bnabla c= 
    D\bnabla^2 c - D \beta  Q E \,\bee\cdot\bnabla \rho + \sqrt{4DC_0} \, \eta_c,  \label{eq:stoch-c} \\
\eeq
and
\beq
\partial_t \rho + \bfv \cdot \bnabla \rho = D\bnabla^2 \rho - D \beta Q E \,\bee \cdot \bnabla c - D \kappa^2 \rho + \sqrt{4DC_0} \, \eta_\rho,\label{eq:stoch-rho-0}
\eeq
where the noise correlations are given as 
$\langle \eta_\rho (\bm{r},t) \eta_\rho (\bm{r}',t') \rangle = \langle \eta_c (\bm{r},t) \eta_c (\bm{r}',t') \rangle = -\bnabla^2 \delta^d (\bm{r}-\bm{r}')  \delta (t-t')$, with $\eta_\rho$ and $\eta_c$ having zero averages and being uncorrelated. The simplification has been used in a variety of different contexts such as a dense population of soft particles~\cite{demery2014generalized}, conductivity of strong electrolytes \cite{demery2016conductivity}, fluctuations of ionic currents across nano-pores \cite{zorkot2018nanopore}, and driven binary mixtures~\cite{poncet2017universal}. The stochastic equation for $\rho$ provides a strong constraint between the two fields in the long time and large length scale limit
\beq
\rho=-\frac{\beta Q E}{\kappa^2} \, \bee \cdot \bnabla c, \label{eq:rho-c}
\eeq
leaving only one soft mode to consider in the effective field theory. In Eqs. \eqref{eq:stoch-c} and \eqref{eq:rho-c} all irrelevant nonlinear terms in the sense of renormalization group theory have been omitted. This procedure can be performed by scaling of these equations according to $\bm{r} \to b \bm{r}$, $t \to b^z t$, $\rho \to b^{\chi_\rho} \rho$, $c \to b^{\chi_c} c$, which yields $z=2$, $\chi_c=-d/2$, and $\chi_\rho=-1-d/2$ for the Gaussian fixed point, followed by calculating the effective scaling exponent for the nonlinear terms and showing that they will be negative in the region of interest (see Refs. \cite{Mahdisoltani2021b,mahdisoltani2021} for details). For the advection term $ \bfv \cdot \bnabla c$ in Eq. \eqref{eq:stoch-c}, this scaling assessment yields $b^{2-d/2}$, which shows that this key nonlinearity is relevant for $d<4$. 

Instead of resorting to standard perturbative renormalization group calculations, here a self-consistent calculation strategy is adopted to study the coupling between density fluctuations in driven electrolytes and the hydrodynamic flow fluctuations. The starting point is to calculate the velocity fluctuations using the linear stochastic theory for the strong electrolytes, which yields
\beqa
&&\aver{v_i(\bfq,\omega) v_j(\bfq',\omega')}= (2\pi)^d \delta^d(\bfq+\bfq') (2\pi) \delta(\omega+\omega') \nonumber \\
&&\hskip0.8cm\times \; \frac{4 D \lae^2}{C_0} (\bqhat \cdot \bee)^2\;\frac{\left[\ee_i-\qhat_i (\bqhat\cdot \bee)\right] \left[\ee_j-\qhat_j (\bqhat\cdot \bee)\right]}{\left[i \omega+D q^2\right]\left[-i \omega+D q^2\right]}.\label{eq:veloc-fluc-2}
\eeqa
The resulting stochastic velocity field $\bfv(\bfr,t)$ is then treated as a correlated noise whose spectrum is given by Eq \eqref{eq:veloc-fluc-2} in the Langevin equation for a tracer particle that follows the stochastic trajectory $\bfr(t)$, namely
\beq
\frac{d}{dt}\bfr(t)=\bfv\left(\bfr(t)\right).\label{eq:Lang-1}
\eeq
This can be used to calculate the MSD as follows
\beq
\Delta L^2(t)=\int_0^t d t_1 \int_0^t d t_2 \, {\cal A}^{\rm L}(t_1,t_2),\label{eq:MSD-AL-def}
\eeq
in terms of the Lagrangian velocity auto-correlation function defined as
\beqa
{\cal A}^{\rm L}(t,t') \equiv \aver{\bfv(\bfr(t),t) \cdot \bfv(\bfr(t'),t')},\label{eq:AL-def-velo-fluc}
\eeqa
which yields
\beqa
&&{\cal A}^{\rm L}(t,t')=\frac{2 \lae^2}{C_0} \int_{\bfq} \aver{e^{i \bfq \cdot [\bfr(t)-\bfr(t')]}} e^{-D q^2 |t-t'|}\nonumber \\
&&\hskip1.4cm\times \; \frac{1}{q^2}(\bqhat \cdot \bee)^2\; \left[1-(\bqhat\cdot \bee)^2\right],\label{eq:velo-fluc-2}
\eeqa
after frequency integration, where we have used the shorthand notation $\int_\bfq \equiv \int \frac{d^d \bfq}{(2 \pi)^d}$. Since the right-hand side of Eq. \eqref{eq:velo-fluc-2} depends on the stochastic trajectory itself, the calculation will need to be closed via a self-consistency requirement. Note that the background linear Langevin noise corresponding to the passive diffusion coefficient ${\cal D}_t$ of the tracer has been ignored in Eq.~\eqref{eq:Lang-1} and throughout this work for simplicity of the presentation. Let us now consider the different regimes that arise from Eqs. \eqref{eq:MSD-AL-def} and \eqref{eq:velo-fluc-2}.

\textit{Velocity fluctuations.---}At the shortest time-scales one can approximate the auto-correlation function in Eq. \eqref{eq:MSD-AL-def} as ${\cal A}^{\rm L}(t,t) \equiv \aver{\bfv(\bfr(t),t) \cdot \bfv(\bfr(t),t)}$, which amounts to using the local and instantaneous velocity fluctuations. These can be calculated as follows
\beq
\aver{\bfv^2}=\frac{2(d-1)}{d (d+2)}\, \frac{\lae^2}{C_0} \,{\cal G}_2^{(d)},\label{eq:v^2}
\eeq
where we have defined the following quantity
\beq
{\cal G}_2^{(d)}=\int_{1/L}^{1/a}\frac{d^d{\bfq}}{(2\pi)^d} \frac{1}{q^2} \sim
    \begin{cases}
        {1}/{a^{d-2}}, & d>2, \\
        \ln\left({L}/{a}\right), & d=2, \\ 
         L^{2-d}, & d<2,
    \end{cases}
\eeq
which reveals strong differences as a function of dimensionality. In particular, it emerges that the ballistic regime defined as $\Delta L^2=\aver{\bfv^2} t^2$ at the shortest time scales is only accessible for $d\geq 2$, as the velocity fluctuations diverge with the system size in $d<2$. 
The anisotropy originating from the symmetry breaking gives rise to different coefficients for the parallel and perpendicular components, namely, $\aver{v_{\parallel}^2}=\frac{3}{d+4}\aver{\bfv^2}$ and $\aver{v_{\perp}^2}=\frac{d+1}{d+4}\aver{\bfv^2}$, as defined by the direction of the electric field $\bee$. Since the anisotropy is expected to only affect numerical prefactors, its effect is not explicitly calculated in the remainder of this Letter for simplicity of the presentation.

\textit{Anomalous diffusion.---}In the intermediate time regime, a self-consistent treatment of the dynamics is needed. To this end, an ansatz of the form $\aver{e^{i \bfq \cdot [\bfr(t)-\bfr(t')]}}=e^{-{\cal D}_z q^z |t-t'|}$ is used for the anomalous diffusion regime in the intermediate time-scales, where $z$ is the dynamic exponent and ${\cal D}_z$ is the corresponding anomalous diffusion coefficient, both of which need to be calculated self-consistently. In the diffusive regime at long times, the form $\aver{e^{i \bfq \cdot [\bfr(t)-\bfr(t')]}}=e^{-{\cal D}_{\rm sc} q^2 |t-t'|}$ will be used, where the effective diffusion coefficient ${\cal D}_{\rm sc}$ is to be calculated self-consistently. Using these forms, the double time integration can be carried out as follows $\int_0^t d t_1 \int_0^t d t_2 \, e^{-\Lambda t}=\frac{2 t}{\Lambda}+\frac{2}{\Lambda^2} \left(e^{-\Lambda t}-1\right)$.

Using $\Delta L^2=\left(2 d {\cal D}_z t\right)^{\alpha}$ for intermediate time scales where $\alpha$ is the anomalous diffusion exponent, which is related to the dynamic exponent $z$ via $\alpha=2/z$, the self-consistency equation reads
\beq
\left({\cal D}_z t\right)^{2/z}=a_d \left(\frac{\lae^2}{C_0}\right)  \int_{\bfq} \frac{e^{-({\cal D}_z q^z+D q^2) t}}{q^2 \left({\cal D}_z q^z+D q^2\right)^2},\label{eq:selfc-1}
\eeq
where $a_d$ is a $d$-dependent numerical prefactor. Note that diffusive contributions have been omitted in Eq. \eqref{eq:selfc-1} as they are sub-dominant with respect to the anomalous super-diffusion in $d<4$. Equation \eqref{eq:selfc-1} gives rise to the emergence of two different scaling regimes, and a crossover between them that occurs at the crossover time 
\beq
\tau^{(d)}_{\textrm{a}\times}=\left(\frac{C_0 D^{d/2}}{\lae^2}\right)^{\frac{2}{4-d}}.\label{eq:tau-aX-def-d}
\eeq
The first scaling regime at relatively shorter times is characterized by 
\beq
z_1=\frac{4}{6-d}, \hskip3mm \alpha_1=3-\frac{d}{2}, \hskip3mm {\cal D}_{z_1}=\left(b_d D^{1-\frac{d}{2}} {\lae^2}/{C_0}\right)^{\frac{2}{6-d}},\label{eq:Dz1}
\eeq
where $b_d=a_d \Gamma\left(-3+{d}/{2}\right)/[2^d \pi^{d/2} \Gamma(d/2)]$.
The derivation of this first scaling regime is equivalent to using the Eulerian velocity autocorrelation function ${\cal A}^{\rm E}(t,t') \equiv \aver{\bfv(\bfr,t) \cdot \bfv(\bfr,t')}$. The second scaling regime at relatively longer times is characterized by 
\beq
z_2=\frac{d}{2}, \hskip3mm \alpha_2=\frac{4}{d}, \hskip3mm {\cal D}_{z_2}=\left(c_d {\lae^2}/{C_0}\right)^{\frac{1}{2}},\label{eq:Dz2}
\eeq
where $c_d=a_d \Gamma\left(-{4}/{d}\right)/[2^{d-2} \pi^{d/2} d\,\Gamma(d/2)]$. The crossover time can be calculated from either of the self-consistent solutions and they turn out to exactly match with one another.

These results are summarized in Table \ref{tab:exponents}. It is interesting to note that $\alpha_1 < \alpha_2$ for $0< d <2$ and $\alpha_1 > \alpha_2$ for $2< d <4$. For $d=2$, we have $\alpha_1=\alpha_2=2$, and for $d\geq 4$, we have $\alpha_1=\alpha_2=1$. Figure \ref{fig:MSD} shows these features (reflected in convexity or concavity of the plots) and the anomalous regimes in different dimensions. 

\begin{table}
    \centering
    \begin{tabular}{ccccc}
    \hline
    \hline
    \hline 
       $d$ & \qquad $z_1=\dfrac{4}{6-d}$ & \qquad $\alpha_1=3-\dfrac{d}{2}$ & \qquad $z_2=\dfrac{d}{2}$ & \qquad $\alpha_2=\dfrac{4}{d}$\\
         \hline
         \hline
      $1$   & \qquad ${4}/{5}$ & \qquad ${5}/{2}$ & \qquad ${1}/{2}$ & \qquad ${4}$\\
      $2$   & \qquad $1$ & \qquad $2$ & \qquad $1$ & \qquad $2$ \\
      $3$   & \qquad ${4}/{3}$ & \qquad ${3}/{2}$ & \qquad ${3}/{2}$ & \qquad ${4}/{3}$\\
      $4$   & \qquad $2$ & \qquad $1$ & \qquad $2$ & \qquad $1$ \\
          \hline
    \hline
    \hline
    \end{tabular}
    \caption{Exponents corresponding to the two types of anomalous dynamics in the intermediate-time regime for different dimensions. Anomalous diffusion ceases to exist for $d>4$.}
    \label{tab:exponents}
\end{table}

\textit{Effective diffusion.---}At long times the system might be able to cross over to a purely diffusive regime. To probe this, a self-consistent equation similar to Eq. \eqref{eq:selfc-1} is used, which yields the following equation for the effective diffusion coefficient
\beq
{\cal D}_{\rm sc}=g_{d}' \, \frac{\lae^2}{C_0} \,{\cal G}_4^{(d)}\,\frac{1}{{\cal D}_{\rm sc}+D},\label{eq:Dsc-def-1}
\eeq
where
\beq
{\cal G}_4^{(d)}=\int_{1/L}^{1/a} \frac{d^d{\bfq}}{(2\pi)^d} \frac{1}{q^4} \sim
    \begin{cases}
        {1}/{a^{d-4}}, & d>4, \\
        \ln\left({L}/{a}\right), & d=4, \\ 
         L^{4-d}, & d<4,
    \end{cases}
\eeq
and $g_d'$ is a $d$-dependent numerical prefactor. Because of the structure of ${\cal G}_4^{(d)}$, the leading order solution to the self-consistency equation given in Eq. \eqref{eq:Dsc-def-1} strongly depends on dimensionality. For $d>4$, ${\cal G}_4^{(d)}$ remains finite and therefore ${\cal D}_{\rm sc} \ll D$, which yields
\beq
{\cal D}_{\rm sc}=\frac{g_{d} \, \lae^2}{D C_0 a^{d-4}}, \hskip10mm (d>4)\label{eq:Dsc-d>4-1}
\eeq
where $g_d$ is a $d$-dependent numerical prefactor. For $d \leq 4$, ${\cal G}_4^{(d)}$ diverges with system size $L$, thereby rendering ${\cal D}_{\rm sc} \gg D$, which yields
\beq
{\cal D}_{\rm sc}=\left(\frac{g_{d} \, \lae^2}{C_0}\right)^{\frac{1}{2}} L^{2-\frac{d}{2}}, \hskip10mm (d<4)\label{eq:Dsc-d<4-1}
\eeq
and
\beq
{\cal D}_{\rm sc}=\left(\frac{g_{4} \, \lae^2}{C_0}\right)^{\frac{1}{2}} \sqrt{\ln(L/a)}. \hskip10mm (d=4)\label{eq:Dsc-d=4-1}
\eeq
Therefore, the diffusive regime is not expected to be accessible for $d<4$ at long times (see below).

\textit{Crossover time-scales.---}It is instructive to examine the behaviour of crossover time-scales for the possible scenarios. The crossover from the ballistic regime to the first anomalous regime occurs at a time-scale $\tau^{(d)}_{\rm ba}$, which can be obtained via $\aver{\bfv^2} (\tau^{(d)}_{\rm ba})^2={\cal D}_{z_1}^{3-d/2} (\tau^{(d)}_{\rm ba})^{3-d/2}$, where we will use input from Eqs. \eqref{eq:v^2} and \eqref{eq:Dz1}. While for $d=2$, this does not yield a result, we find $\tau^{(d)}_{\rm ba}=a^2/D$ for $d>2$ and $\tau^{(d)}_{\rm ba}=L^2/D$ for $d<2$. This implies that the ballistic regime only exits for $d>2$, as the velocity fluctuations diverge with the system size in $d<2$ (see above). 

The crossover time from ballistic regime to diffusive regime, $\tau^{(d)}_{\rm bd}$, can be obtained via $\aver{\bfv^2} (\tau^{(d)}_{\rm bd})^2={\cal D}_{\rm sc} \tau^{(d)}_{\rm bd}$. Note that this crossover exists only for $d \geq 4$. Using Eqs. \eqref{eq:v^2}, \eqref{eq:Dsc-d>4-1}, and \eqref{eq:Dsc-d=4-1}, the calculation yields $\tau^{(d)}_{\textrm{b}\textrm{d}}=C_0^{1/2} \lae^{-1} a^{d/2}$ for $d>4$ and $\tau^{(4)}_{\textrm{b}\textrm{d}}=C_0^{1/2} \lae^{-1} a^2 \sqrt{\ln(L/a)}$ for $d=4$. For $d<4$, a long-time crossover occurs at time $\tau_{L}^{(d)}$ when the MSD in the second anomalous regime is saturated by the system size $L$. Using $L^2={\cal D}_{z_2}^{4/d}(\tau_{L} ^{(d)})^{4/d}$ and Eq. \eqref{eq:Dz2}, the calculation gives $\tau_{L}^{(d)}=C_0^{1/2} \lae^{-1} L^{d/2}$ for $d<4$. Note that a similar calculation using the divergent self-consistent effective diffusion coefficient from Eq. \eqref{eq:Dsc-d<4-1} yields the same result, namely, $\tau_{L}^{(d)}=L^2/{\cal D}_{\rm sc}=C_0^{1/2} \lae^{-1} L^{d/2}$ for $d<4$. Finally, since in $d=2$, the anomalous diffusion exponents for both regimes are the same as the ballistic dynamics exponent, i.e. $\alpha_1=\alpha_2=2$, the ballistic regime is expected to govern all time scales in this case (see Fig. \ref{fig:MSD}b). Note that the adiabatic elimination of the charge density due to Debye screening relies on the Debye relaxation time $1/(D \kappa^2)$ being much shorter than the other relevant timescales such as $1/\lae$ or the various crossover scales. This is guaranteed by the strong electrolyte condition, as can be seen for example by noting that for physiologically relevant salt concentrations $\kappa^{-1} \sim 1$ nm giving rise to a Debye relaxation time $\sim 1$ ns.

\textit{Concluding remarks.---}The stochastic fluctuations of tracers in a bulk electrolyte have been found to exhibit a wide range of dynamical regimes and crossover scales that depend strongly on dimensionality. The existence of two anomalous regimes is particularly interesting, as they correspond to a crossover between a regime where density fluctuations directly result in anomalous dynamics (characterized by $\alpha_1=3-d/2$) to a regime where advection by the fluid flow changes the nature of the anomalous dynamics (characterized by $\alpha_2=4/d$) albeit driven by the same density fluctuations. This phenomenon is intrinsically related to the anomalous dynamics of catalytically active colloids where only the first regime was originally considered \cite{Golestanian2009b,Golestanian2019phoretic}, due to similarities in the interplay between density fluctuations and dynamics under non-equilibrium driving forces and fluxes. Anomalous dynamics can also arise in anisotropic active suspensions \cite{sriram0}. Because of the important role that ionic currents play in molecular sensing technologies \cite{Ying2022,Ritmejeris2024}, it will be important to investigate how such strong and anomalous fluctuations can influence these sensing devices. A starting point can be to incorporate hydrodynamic and ionic effects in simple models of molecular sensing that use polymer translocation \cite{Cohen2012}.

\acknowledgements
The author acknowledges discussions with Sriram Ramaswamy and support from the Max Planck School Matter to Life and the MaxSynBio Consortium which are jointly funded by the Federal Ministry of Education and Research (BMBF) of Germany and the Max Planck Society.

\end{document}